\documentclass[preprint2]{proto}
\usepackage{times}
\newcommand{\refs}{\par\noindent\hangindent=1pc\hangafter=1}
\begin{document}

\title{\textbf{\LARGE Disk Evolution in Young Binaries: from Observations to Theory}}

\author {\textbf{\large J.-L. Monin} }
\affil{\textbf{Laboratoire d'Astrophysique de Grenoble, France}}

\author {\textbf{C. J. Clarke}}
\affil{\textbf{Institute of Astronomy, Madingley Road, Cambridge, CB3 0HA, UK}}

\author {\textbf{L. Prato}}
\affil{\textbf{ Lowell Observatory, 1400 West Mars Hill Road,
Flagstaff, AZ 86001, USA}}

\author {\textbf{C. McCabe}}
\affil{\textbf{JPL, 4800 Oak Grove Drive, Pasadena, CA 91109 USA}}

\bigskip

\begin{abstract}
\baselineskip = 11pt
\leftskip = 0.65in 
\rightskip = 0.65in
\parindent=1pc

{\small 
The formation of a binary system surrounded by disks is the most common
outcome of stellar formation. Hence studying and understanding the
formation and the evolution of binary systems and associated disks is a
cornerstone of star formation science. Moreover,  since
the components within binary systems are coeval and the sizes of
their disks are fixed by the tidal truncation of their companion,
binary systems provide an ideal "laboratory" in which to study
disk evolution under well defined boundary conditions.

Since the previous edition of Protostars and Planets, large diameter
(8$-$10m) telescopes have been optimized and equipped with adaptive
optics systems, providing diffraction-limited observations in the
near-infrared where most of the emission of the disks can be traced.
These cutting edge facilities provide observations of the inner parts
of circumstellar and circumbinary disks in binary systems with
unprecedented detail.  It is therefore a timely exercise to review the
observational results of the last five years and to attempt to
interpret them in a theoretical framework.

In this paper, we review observations of several inner disk
diagnostics in multiple systems, including hydrogen emission
lines (indicative of ongoing accretion), $K-L$ and $K-N$ color
excesses (evidence of warm inner disks), 
and polarization (indicative of the relative orientations of the
disks around each component). We examine to what
degree these properties are correlated within binary systems and
how this degree of correlation depends on parameters such as
separation and binary mass ratio. These findings will be interpreted
both in terms of models that treat each disk as an isolated reservoir
and those in which the disks are subject to re-supply from some form of
circumbinary reservoir, the observational evidence for which we will
also critically review. The planet forming potential of multiple star
systems is discussed in terms of the relative
lifetimes of disks around single stars, binary primaries and binary
secondaries.  Finally, we summarize several potentially revealing 
observational problems and future projects that could provide further
insight into disk evolution in the coming decade.\\
%
 \\~\\~\\~}
\end{abstract}  

\section{\textbf{INTRODUCTION}}

  It is now a matter of common knowledge that the majority of
stars in star forming regions are in binary or higher order multiple
systems ({\em Ghez et al.}, 1993; {\em Leinert et al.}, 1993; {\em Simon et al.}, 1995).
Likewise, it is undisputed
that many of the younger stars in these regions exhibit evidence
for circumstellar disks and/or accretion. Putting these two facts together,
an inescapable conclusion is that disks typically form and
evolve in the environment of a binary/multiple star system.

  This prompts a number of obvious questions. Can the distribution
of dust and gas in young binaries provide a ``smoking gun'' for the
binary formation process? Is disk evolution, and perhaps the possible 
formation of planets, radically affected by the binary environment and, if so,
how does this depend on binary separation and mass ratio? Alternatively,
if the influence of binarity on disk evolution is rather mild, we can
at least use binary systems as well controlled laboratories, constituting
coeval stars with disk outer radii set by tidal truncation criteria,
to  study    disk evolution as a function of stellar mass.

  However, it is not possible to address any of these issues unless we can
disentangle the disk/accretion signatures produced by each component 
in the binary. Given that the separation distribution for binaries
in the nearest populous star forming regions, such as Taurus-Aurigae,
peaks at $\sim0.3''$ ($\equiv 40\,$AU; e.g., {\em Mathieu}, 1994), this 
necessitates the use of high resolution photometry and spectroscopy.
Such an enterprise has only become possible in the past decade.

   We review what has been learned in recent years about
the distribution of dust and gas within  young binary systems.
We mainly highlight observational developments since PP\,IV, for
example, the discovery of a population of so called passive disks
({\em McCabe et al.}, 2006) in low mass secondaries and the use of polarimetry
to constrain the orientations of disks in young binaries (e.g., {\em Jensen
et al.}, 2004; {\em Monin et al.}, 2005). We also discuss circumbinary disks
and profile in detail a few systems that
have been the subject of intense observational scrutiny. In addition,
it is timely to examine the statistical
properties of resolved binaries that have been accumulating in the
literature over the past decade. We have therefore combined
the results from a number of relatively small scale studies in order to
assemble around $60$ resolved pairs and use this dataset to
examine the relationship between binarity and disk evolution.

In this Chapter we progress through a description of disk/accretion
{\it diagnostics} (and their application to resolved binary star studies;
Section~\ref{sec:diag}) to highlighting some
recent results on disk {\it structure}
in binaries (Section~\ref{sec:struc}) to a statistical analysis of the
relationship between binarity and
disk {\it evolution} (Section~\ref{sec:evol}).
In Section~\ref{sub:planets} we briefly consider
how the insights of the preceding
section can be applied to the question of planet formation in binaries.
Section~\ref{sec:future} examines future prospects and potential projects
to advance our understanding of disk structure and evolution in young binaries.

\section{INNER DISK DIAGNOSTICS IN YOUNG BINARIES}
\label{sec:diag}

How do we know when either circumstellar or circumbinary disks are present
in a young multiple star system?  It took over a decade of
observations to confirm the existence of simple circumstellar disk
structures after the original observational and theoretical introduction
of the concept in the early 1970s (e.g., {\em Strom et al.}, 1971, 1972;
{\em Lynden-Bell and Pringle}, 1974).  The paradigm is yet more complicated 
for a binary systems with multiple disks. 
Over the last two decades, direct means of imaging circumstellar
disks have become available to astronomers, beginning with millimeter
observations in the mid-1980s and ending with the recent development
of high angular resolution laser guide star adaptive optics.  Most
critically to this chapter, the last decade has witnessed
unprecedented improvements in our ability not only to directly image
disks and to indirectly infer their presence, but to detect
disks around both stellar components of extremely close binary
systems, as well as larger, circumbinary structures.

In this section we summarize the methods used to determine the
presence of circumstellar and circumbinary disks in multiples.
The section is divided into two parts:
disk diagnostics  and accretion diagnostics.  In this manner we
distinguish between observations that detect the disks themselves,
directly or indirectly, and the observations that are sensitive to
the presence of accretion processes, indicating that material is
flowing from a disk onto the central star, and thus betraying
the existence of the disk indirectly.  The end of this section summarizes
the database that we compiled in the process of writing this chapter.

An excellent inner disk diagnostic that we do not explore is the 
emission of molecular lines.  This topic is reviewed in the PP\,V paper
of {\em Najita et al.} as well as in {\em Najita et al.} (2003) and references therein.
Outer disk diagnostics, such as sub-millimeter and millimeter observations,
and more narrowly applied accretion diagnostics, such as forbidden line
emission and ultraviolet excesses, are also neglected here because they
do not appear in our analysis.  These tools are either not as relevant to our
component resolved studies or have not yet been widely applied to many
binary observations.  Relevant references may be found in {\em Dutrey et al.} (1996)
and {\em Jensen et al.} (1996) (sub-millimeter and millimeter), {\em Hartigan and
Kenyon} (2003) ([OI] emission lines), and {\em Gullbring et al.} (1998).

\subsection{Background}

{\em Lynden-Bell and Pringle}'s (1974) prescient disk model for
classical T Tauri stars (CTTs) accounted for a number of
characteristics of these objects, including atomic emission
lines and the relatively
flat $\lambda F_{\lambda}$ distribution of light at infrared
wavelengths.  Pioneering infrared observations of young stars
indicated the presence of a strong excess
({\em Mendoza} ,1966, 1968)
above expected photospheric values ({\em Johnson}, 1966) for T Tauri stars.
These were largely interpreted as indicative of either a
spherical dust shell around the young stars studied
({\em Strom et al.}, 1971; {\em Strom}, 1972; {\em Strom et al.}, 1972) or
free-free emission from circumstellar gaseous envelopes
({\em Breger and Dyke}, 1972; {\em Strom et al.}, 1975), although the suggestion
of a circumstellar disk structure was raised as early as 1971
({\em Strom et al.}, 1971, 1972).  Early analysis of the IRAS satellite data (e.g., {\em Rucinski}, 
1985) and direct imaging of disk-like structures around HL Tau
and L 1551 IRS 5 ({\em Grasdalen et al.}, 1984; {\em Beckwith et al.}, 1984;
{\em Strom et al.}, 1985) provided ultimately convincing evidence in
support of disks.

\subsection{Disk Diagnostics}
\subsubsection{Near- and Mid-Infrared Excesses}

\begin{figure}[htb]
 \epsscale{1.0}
\plotone{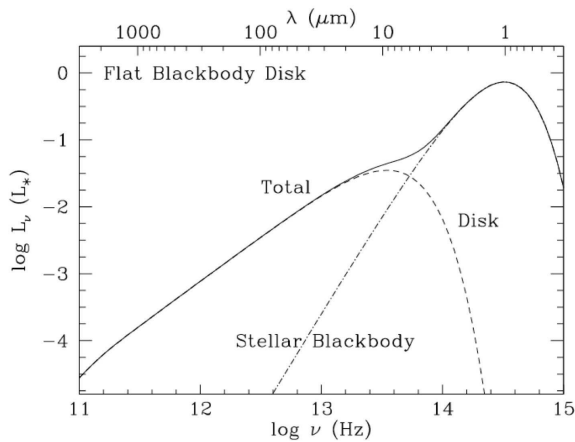}
 \caption{SED of a flat reprocessing disk from {\em Chiang and Goldreich} (1997).
The dashed line corresponds to the disk emission, the dot-dash line to the stellar
photosphere, and the solid line shows the total flux. \label{fig:sd-sed}}  
 \end{figure}

Optically thick but physically thin dusty circumstellar disks
around T Tauri stars reprocess stellar flux and
give rise to excess thermal radiation at wavelengths greater
than $\sim\,1\,$micron. 
At larger disk radii, the equilibrium dust temperature is lower;
thus, different circumstellar disk regions are studied
in different wavelength regimes.  For low-mass stars, the $JHK (1-2\,\mu{\rm m})$
colors sample the inner few tenths of an AU, the $L$-band ($3.5\,\mu{\rm m}$)
about twice that distance, the $N$-band ($10\,\mu{\rm m}$) the inner $\sim$1$-$2~AU.
$IRAS$ and $Spitzer$ data sample radii of several to tens of AU.
The exact correspondences depend on the luminosity of the central star
and the disk properties and geometry (e.g., scale height of the dust,
degree of flaring, particle size distribution; see e.g., {\em D'Alessio} 2003;
{\em Chiang et al.} 2001; {\em Malbet et al.} 2001). 
This results in a spectral energy distribution where the disk and star contributions can be disentangled (Fig.~\ref{fig:sd-sed}). 
For the nearby star forming regions, binary separation
distributions typically peak at $\sim\,0.3''$ ({\em Simon et al.} ,1995;
{\em Patience et al.}, 2002); for a
distance of 150 pc, this corresponds to 45 AU.  Therefore,
most stars in binaries should have a direct impact on the circumstellar
environments of their companions,
at least at radii of several to a few dozen AU from the individual stars.

The shortest wavelength $JHK$ colors, although useful
(e.g., {\em Hillenbrand}, 1998),
are susceptible to contamination from reflected light and are
highly sensitive to the disk inner gap size
(e.g., {\em Haisch et al.}, 2000). $L$-band data, where the contribution from the T Tauri stellar photosphere decreases, offers a far more
reliable indicator of the innermost
circumstellar dust (e.g., {\em Haisch et al.}) and
reveals a much larger proportion of stars with disks.
In the 1993 proceedings from Protostars and Planets III,
{\em Edwards et al.} summarized the relationships between the
K$-$L disk colors and the winds off of, and accretion flows on to,
T Tauri stars.  This establishment of the usefulness of
K$-$L colors as a disk diagnostic coincided with early studies
of binary colors ({\em Leinert and Haas}, 1989; {\em Ghez et al.}, 1991).  Since
the mid-1990s this diagnostic has been widely used as a convenient
indicator of circumstellar disks in small separation binaries
(e.g., {\em Tessier et al.}, 1994; {\em Chelli et al.}, 1995; {\em Geoffray and Monin}, 2001;
{\em White and Ghez}, 2001; {\em Prato et al.}, 2003; {\em McCabe et al.}, 2006).
At 8$-$10 m class telescopes, an angular resolution of $\sim$0.1$''$ is
achievable in the $L$-band.

$N$-band observations are sensitive to dusty disk material
that may surround a young star even in the absence of an
innermost disk and a corresponding near-infrared excess
(e.g., {\em Koerner et al.}, 2000; {\em Prato et al.}, 2001).
For more than 30 years, observations have been made of young
stars at 10 and $20\,\mu$m (e.g., {\em Strom et al.},  1972;
{\em Knacke et al.}, 1973; {\em Strom et al.}, 1975; {\em Rydgren et al.}, 1976;
{\em Skrutskie et al.}, 1990; {\em Stassun et al.},  2001), but,
with few exceptions (e.g., {\em Ghez et al., 1994}) high angular
resolution mid-infrared observations required the development
of a new generation of cameras in the late 1990's for the largest existing
(8$-$10 m class) telescopes.  The Keck
telescopes, for example, provide a 0.25$''$ diffraction limit
at $10\,\mu$m. Over 80\%
of the known, angularly resolved, young binary $N$-band measurements,
and most of the angularly resolved $Q$-band ($\sim 20\,\mu$m)
measurements, have only recently been published in {\em McCabe et al.} (2006).
Although far-infrared space-based observations do not provide the 
requisite angular resolution to distinguish between close
binary components ($Spitzer's$ diffraction limit at $160\,\mu$m
is about half a minute of arc), ALMA will provide unprecedented
resolution in the far-infrared/sub-millimeter regime (see section~\ref{sec:future}).

\subsubsection{Polarization}

Linear polarization maps of young stars typically show an
axisymmetric, or "centrosymmetric" pattern.  By the late 1980's, these
observations were interpreted by {\em Bastien and M\'enard} (1988, 
1990) as the result, in part, of light scattering from optically thick
circumstellar disks.  A prescient remark from the PP\,III paper of
{\em Basri and Bertout} (1993) notes that "High resolution
near-infrared polarization maps are, however, becoming possible with
the advent of 256x256 detectors and AO..."  Indeed, by the late
1990's, stunning detail in the polarization maps of {\em Close et al.}
(1998), {\em Potter et al.} (2000), and {\em Kuhn et al.} (2001)
illustrated the power of polarization observations for the study of
circumstellar and circumbinary disks.  {\em Monin et al.} (1998)
applied the tool of polarization to a sample of wide (8$-$40$''$)
binaries in Taurus.  Most recently, {\em Jensen et al.} (2004) and
{\em Monin et al.} (2005) mapped polarization around more than 3 dozen
small separation ($\sim$1$-$10$''$) binaries (\S 3.1).  Given that
polarization observations can identify the orientation of a
circumstellar disk, this provides a unique way in which to test the
alignment of disks in binary systems.

\subsection{Accretion Diagnostics: Permitted atomic line emission}

The prolific work in the 1940's of Joy (e.g., {\em Joy and van
Biesbroeck}, 1944) and later of Herbig (e.g., {\em Herbig}, 1948) on T
Tauri type emission line stars laid the foundations for the study of
emission lines in young binaries.  Although the source of hydrogen
emission lines in young stars was variously attributed as the result
of free-free emission, chromospheric activity, and stellar winds
(e.g., {\em Strom et al.}, 1975; {\em Herbig}, 1989; {\em Edwards et
al.}, 1987), by the late 1980's {\em Bertout et al.} (1988) and others
had established a model for magnetospheric accretion.  {\em Strom et
al.} (1989) canonized the nominal 10\,\AA~ distinction in H$\alpha$ ($\lambda=6563\,$\AA)
line emission between weak-lined (wTT) and classical (cTT) T Tauri
stars, which is still - somewhat indiscriminately - used today, albeit
with slight modifications (e.g., {\em Mart\'{\i}n}, 1998).

As the high frequency of young star binaries became established in the
mid-1990's, hydrogren emission lines were recognized as a useful
approach for the study of circumstellar material around each star in
the system (e.g., {\em Hartigan et al.}, 1994; {\em Brandner and
Zinnecker}, 1997; {\em Prato and Simon}, 1997; {\em Duch\^ene et al.},
1999; {\em White and Ghez}, 2001; {\em Prato et al.}, 2003).  Infrared
Br$\gamma$ ($\lambda=2.16\,\mu$m) observations ({\em Prato and Simon}, 
1997; {\em Muzerolle et al.}, 1998; {\em Prato et al., 2003}) provide a
means of measuring emission lines with the best possible seeing at
longer wavelengths, as well as a method of detecting infrared
components not readily seen in visible light.

\subsection{The Young Star Binary Database}

Observations of young binaries that resolve the circumstellar disk and
accretion diagnostics of each component involve access to large
telescopes with adaptive optics capabilities or space based
observatories.  Limited access to such facilities has meant that the
results of such studies have often been published in papers describing
a relatively small number of objects, from which it has proved
impossible to derive statistically secure results.  We have therefore
combined many studies into a single database.  In order to qualify for
inclusion in the database, it is only necessary for the binary
components to be angularly resolved and located in a region with a
distance estimate such that the separation of the pair in the plane of
the sky is known.  Because we restrict the database to resolved
systems, we exclude systems with semi-major axis $a < 14$~AU.  In
order to avoid contamination with chance projections we also exclude
systems with $a > 1400$~AU. These limits correspond to binaries in the
angular separation range of $0.1-10.0''$ at the 140~pc distance of
Taurus.

Although the information available is incomplete for a number of
objects, we have $\sim$60 systems where the spectral type of each
component is known, as well as the presence or absence of disks and/or
accretion for each component.  We shall return to the statistical
properties of these systems, and the implications for disk evolution
in binaries, in \S 4.  The database is available at
http://www.lowell.edu/users/lprato/compil\_binaires\_cmc5.html.  We
welcome additions, revisions, and comments.

\section{DISK STRUCTURE IN YOUNG BINARIES}
\label{sec:struc}

\subsection{Disk orientations in binary systems}

A binary system with disks possesses many more degrees of freedom than
an isolated star. Both stars can have a disk, they orbit around each
other, and the entire system can be surrounded by a circumbinary disk.
This defines 4 planes: 2 circumstellar disks, the stellar binary
orbit, and the circumbinary disk.  In a single young star system, only
one plane is potentially present, that of one disk.  In this section
we examine some recent observational and theoretical results that shed
light on the respective orientations of the multiple planes present in
a binary.

If a binary forms through the fragmentation of a disk, then the disks
that form around each star are expected to be mutually aligned and
also to be aligned with the binary orbit. The same is true for binaries
that form through core fragmentation, 
provided the angular momentum
vectors of the parent core material are well aligned and provided
that the initial core geometry (or the result of any perturbation inducing 
fragmentation) does not introduce any other symmetry planes into
the problem. Although {\em Papaloizou and Terquem} (1995) showed that tidal
effects may sometimes 
induce subsequent misalignment of the disks
and orbital planes, {\em Lubow and Ogilvie} (2000) show that the required conditions
are unlikely to be met in practice. We shall therefore assume that
a binary that is created with all its planes aligned will remain
in this state throughout its Class 0, Class I and Class II phases.

If any of the above conditions are not met, however, the binary will
be created with some planes misaligned. For example,
{\em Bonnell et al.} (1992) showed that if the initial cloud is elongated and
if the rotation axis is oriented arbitrarily with
respect to the cloud axis, then the initial disk and binary orbital
planes are misaligned: in this case, the disk planes (which reflect
the angular momentum of the core) are parallel, and misaligned with
the binary orbit (which reflects the symmetry of the initial core).
On the other hand, all planes
may be misaligned either in the case that the angular momentum
distribution of the initial core is complex or that the fragmentation
involves more than two bodies. 
There are therefore a number
of routes by which misaligned systems can be created and may be manifest
among Class 0 systems. 
This does not, however, imply that
these systems will remain misaligned during subsequent evolutionary
phases, owing to the fact that both tidal effects and
accretion onto the protobinary can bring the system into
alignment at a later stage. Therefore the detection of misaligned
systems is an unambiguous sign of misaligned formation, whereas
aligned systems may either have been created that way or else
have subsequently evolved into this state.  

At the earliest evolutionary stages, it now  seems inescapable 
that at least some systems contain misaligned disks. In these
systems, jet orientation provides an observable proxy for disk
orientation since jets are always launched perpendicular to
the inner disk: the detection of multiple jets emanating
with different position angles from a small region is thus an unambiguous
sign of misalignment ({\em Reipurth et al.},
1993; {\em Gredel and Reipurth}, 1993; {\em Davis et al.}, 1994; {\em Bohm and Solf}, 1994;
{\em Eisloffel et al.}, 1996).  In all cases, the parent multiple
systems are either unresolved or are known to be wide binaries
(i.e. with $a > 100$~AU). Less directly, the
observation of changes in jet position angle have been 
interpreted as the result of jet precession (or ``wobble''; {\em Bate et al.},
2000), induced by misalignment between the disk and the orbital plane
of a putative companion ({\em Chandler et al.}, 2005; {\em Hodapp et al.}, 2005).
The observed rates of change of jet position angle are thought to be
consistent with the presence of unresolved binary companions with
separations in the range several to $\sim100$~AU. However, 
not all observed instances of changes in jet direction
can necessarily be explained in these terms ({\em Eisloffel and
Mundt}, 1997).

The expected timescale on which strongly misaligned
disks should be brought into rough alignment by tidal torques is
about $20$ binary orbital periods ({\em Bate et al.}, 2000); it is  thus
only in  rather wide binaries (i.e. with $a > 100$~AU) that we should
expect misaligment throughout their Class 0 and Class I stages.
However, as the misalignment
angle ($\delta$) evolves towards zero, the rate of alignment becomes
proportional to $\delta$ and hence the system may be expected to
remain in a mildly misaligned state over considerably longer
periods.

How are these expectations borne out by observations of
Class II sources (with typical ages of a few $\times  10^6$ years)?
The most obvious approach is through direct imaging of disks in PMS binaries. 
Unfortunately, despite the recent deployment of a range of  instruments
offering  high angular resolution on very large telescopes, circumstellar
disks in TTS multiple systems have only been
imaged in few cases (HK Tau: {\em Stapelfeldt et al.},
1998; HV Tau: {\em Monin and Bouvier},  2000; {\em Stapelfeldt et al.},  2003;
LkH$\alpha$~263: {\em Jayawardhana et al.}, 
2002). In each of these systems only one disk is detectable via imaging and is
seen edge-on, a favorable orientation for detection.  In all three systems, the observed edge-on 
disk is oriented in a direction quite different from the projection of
the binary orbit on the sky:
therefore we see immediately that {\it at least some disks in binaries
remain misaligned with the binary orbital plane during the Class II phase}. 

Several properties of these imaged systems
are noteworthy: first, they are all wide binaries 
($a > $ several hundred AU) and are thus consistent with the
estimate given above that disks in binaries closer than $\sim100$~AU 
should be brought into alignment during the Class 0 or Class I phase. Second,
for HK~Tau and LkH$\alpha$~263, the companion to the star with the edge on disk
is itself a close binary system. 
Third, in each of these systems only
one disk is detectable through imaging, although there is some spectroscopic evidence that the other component does possess a disk. 
The fact that these other disks are not
detected through direct imaging implies that they are not themselves viewed
close to edge on and we can thus infer that the disks in these
systems are not parallel with each other. However, 
since only a slight tilt of the other disk away from edge-on
can abruptly reduce its detectability as the central star becomes
visible directly, this observation only
excludes an alignment between the disks  to within
$\approx$ 15\,degrees.

Since the publication of PP\,IV, various studies have been performed to
determine the orientation of binary disks relative to each other in the plane
of the sky. Following the theoretical computations by {\em Bastien and M\'enard} (1990),
and the previous measurements of {\em Monin et al.} (1998), {\em Wolf et al.} (2001),
{\em Jensen et al.} (2004), and {\em Monin et al.} (2005) have 
used polarimetric observations to determine the relative orientation of
disks in the plane of the sky. The position angle of the integrated linear
polarization of the scattered starlight is parallel to the equatorial
plane of the disk, provided that its inclination is sufficiently large
to mask the direct light from the star ({\em Monin et al.}, 2005). 
One caveat of this method is  that it does
not reveal the actual 3D orientations of disks; two disks with parallel
polarization could be differently inclined along the line of sight.
In principle, this other orientation angle can be obtained from $v\sin i$ and 
rotation period measurements,  but these are 
quite rare and difficult to obtain in close binaries  and are in
any case subject to errors attributable to uncertainties in the stellar radius.
However, {\em Wolf et al.} (2001) have shown from statistical arguments that if
the relative polarization position angle difference distribution peaks at
zero, then the disks tend to be parallel. 

\begin{figure}[htb]
 \epsscale{1.0}
\plotone{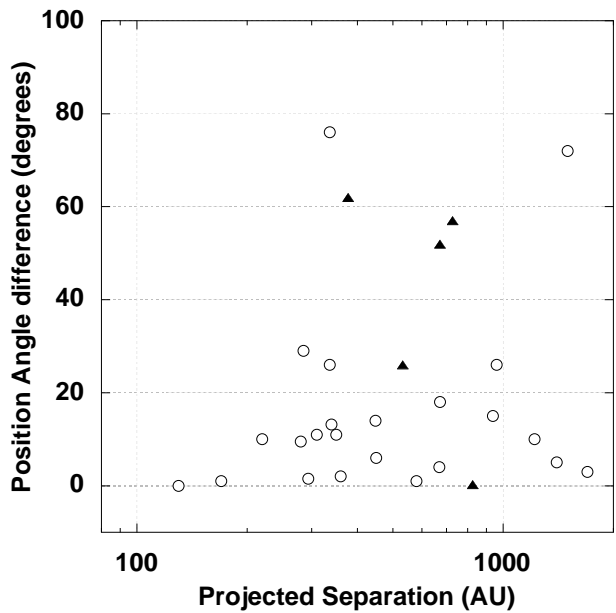}
 \caption{Adapted from {\em Jensen et al.} (2004) and {\em Monin et al.} (2005); binaries are plotted as
empty circles and higher order multiples as filled triangles. Note the suggestion in the
data that disks in triples and quadruples are proportionally  less aligned
than in pure binary systems.
 \label{fig:j+04}} 
 \end{figure}






The net result of these studies is twofold: disk polarizations tend to be 
close to (but not exactly) parallel in binary systems, but there exist systems
with misaligned polarization, with a few objects having polarization position
angle differences of $\sim90^{o}$.  
{\em Jensen et al.} (2004) argue that it is unlikely that
this result is  compromised by dilution of the polarization
signal from each disk 
by interstellar polarization, since, among other evidence, they note
that disk polarization tends {\it not} to be parallel
in the case that one or other of the two components is itself
a close  binary system (see Fig.~\ref{fig:j+04} where we have merged the results
from {\em Jensen et al.} (2004) and {\em Monin et al.} (2005)). 
This supports the notion that the
polarization differences are intrinsic, since there should be no 
correlation between
the degree of contamination by interstellar polarization and
the multiplicity of the system studied.
On the other hand, there is a plausible physical
reason for this result: owing to the much larger angular
momentum contained in a close binary pair than a simple star-disk
system,  the timescale for torquing
a binary into alignment with the orbital plane of the wider pair
is evidently much longer than that for the alignment of a   disk.

 All of the above discussion relates to binaries that are wide
enough to be imaged (typically wider than $100$~AU).
However, in the case of spectroscopic binaries, there is the
possibility of determining the system inclination from the
orbital solution and then comparing this with the inclination
determined from direct imaging of the circumbinary disk
(albeit on a much larger scale). In the small number of
systems where this has proved possible, the evidence is for
alignment between the plane of the spectroscopic binary and its
circumbinary disk (see {\em Mathieu et al.}, 1997; {\em Prato et al.}, 2001;
{\em Prato et al.}, 2002; {\em Simon et al.}, 2000).  


  Finally, among {\it main-sequence} solar type binaries, it is found that
the stellar orbital planes are aligned with the binary orbit for
binaries closer than $40$~AU (Hale 1994), as one
would expect, given the short predicted alignment timescales
for closer binaries.

 In summary, we have plenty of examples (through imaging
and polarization studies) of misaligned systems among wider
binaries (i.e. with $a > 100$~AU or so), implying that at least some
of these systems must be formed in a misaligned state. By the Class II
phase, it would appear that binaries in this separation range
constitute a mixture of aligned and misaligned systems ({\em Monin et al.}, 2005,
{\em Jensen et al.}, 2004). This may  imply that wider binaries
are formed in both aligned  and misaligned states, or, alternatively,
that  all such binaries are born in the misaligned
state and are brought into alignment through tidal torques (which
should operate on a roughly $10^6$ year timescale for binaries of 
this separation).
In the case of closer binaries, where direct imaging is
not possible,   observational 
evidence for disk alignment can be derived only in the
case of spectroscopic binaries with imaged circumbinary disks
and also through the fossil evidence
contained in stellar spin vectors within  main-sequence binaries.
Both these lines of evidence point to 
close  binaries being aligned during the
main disk accretion stage. This is expected, given the short predicted
alignment timescales for close binaries, and therefore gives
us no information about the {\it initial} state of alignment
of these systems.

\subsection{A sampling of circumbinary disks}


Only a few circumbinary disks have been imaged
directly. {\em M\'enard et al.} (1993) proposed a circumbinary disk to
explain NIR images of Haro~6-10, and in 1994, the circumbinary disk that still today
remains the most impressive to date was found by {\em Dutrey et al.} (1994)
around GG~Tau. 
The majority of the currently inferred circumbinary disks are proposed
to explain SED emission from warm dust in disks with a central hole
where the binary resides. With the ever growing number of discoveries
of PMS spectroscopic binaries, the number of putative circumbinary
disks in these closer systems has increased.  On the other hand, in
the case of wide binaries, very few circumbinary disks have been
directly imaged and, moreover, the low upper limits for circumbinary
disk masses from millimetre measurements ({\em Jensen et al.}, 1996;
see also \S 4.4) suggest that circumbinary disks are weak or absent in
the majority of these systems. However, this conclusion remains
provisional on two grounds.  First, the very small number of
circumbinary disks that have been imaged might not be as surprising as
originally thought, when one considers also the relatively low rate of
detection of circumstellar disks by direct imaging; only when the
system geometry is very favorable can the disk be imaged easily (see
section~3.1 above).
Second, there is at least one case in which a circumbinary disk that
has been imaged in CO lines is not detectable in dust as probed by the
millimetre continuum (see discussion of SR 24 N below). We therefore
cannot rule out that wide binaries either possess low mass disks that
escape detection in the dust continuum (corresponding to disk masses
$\le$ a Jupiter mass) or else that some process, such as grain growth,
is reducing the dust emission in these systems. Such a process may be
at work in the GG~Tau circumbinary disk (see \S\,\ref{subsub:ggtau}
below).

In the case of wider binaries ($a > 20$~AU), the argument in favour of
circumbinary disks as a necessary reservoir for the resupply of
circumstellar disks has weakened since its orginal proposal by {\em
Prato and Simon} (1997): our analysis described in \S 4 below shows
that mixed systems (i.e. pairs containing both a cTT and a wTT) are in
fact common.  It is likely that, in wider binaries, circumstellar
disks evolve in relative isolation, and resupply might not be a
necessity. In closer binaries, resupply remains a necessity on the
grounds that the circumstellar disk lifetimes in these close systems
would otherwise be too short to explain the incidence of component cTT
stars. In these closer systems, circumbinary disks, as evidenced by
their contribution to the spectral energy distribution, remain a good
candidate for the resupply reservoir. Indeed, in various objects,
signatures of accretion episodes from the circumbinary environment
onto the central objects, presumably via their associated
circumstellar disks, have been detected.  In this section we examine
in more detail several circumbinary disk systems and discuss their
properties in terms of disk evolution, circumbinary accretion, and
potential for planet formation.

\subsubsection{GG Tau}
\label{subsub:ggtau}

Discovered by {\em Simon \& Guilloteau} (1992), this circumbinary disk
orbits the 0.25$''$ separation pair GG\,Tau~A and has been spatially resolved in
the optical, (Krist et al. 2002; 2005), near-infrared ({\em Roddier et
al.}, 1996; {\em McCabe et al.}, 2002; {\em Duch\^ene et al., 2004}) and
in the millimeter (continuum and $^{12}$CO, e.g., {\em Guilloteau et
al.}, 1999). {\em Beust and Dutrey} (2005) investigated the GG\,Tau~A
orbit and the inner ring gap and find that a binary orbital solution
with $a=62\,$AU and $e=0.35$ could be consistent with the data; in
this study, the presence of the circumbinary disk is used to add
dynamical constraints to the central binary system.  Using a
collection of images at various wavelengths, {\em Duch\^ene et al.}
(2004) have shown that grain growth is at work in the midplane of the
GG~Tau circumbinary ring within a stratified structure. This shows
that the processes leading to planet formation might be at work in
circumbinary disks as well as in circumstellar disks.



\subsubsection{SR24N}
The binary separation in this system is of the same order as  GG~Tau's,
$32\,$~AU.  {\em Andrews and Williams} (2005) have observed a 250~AU structure in
this system, probably a circumbinary disk.  An interesting feature of their
observations is that this disk shows no emission in the continuum, possibly as
the result of a central gap inside the disk, and is seen only in CO line
emission. This suggests that other wide circumbinary disks could have been
missed by continuum observations, and thus could be more frequent than
previously thought. K-L measurements by {\em McCabe et al.} (2006) indicate that
both components of SR24N are themselves cTT stars. 

\subsubsection{GW Ori}
GW~Ori is a spectroscopic binary with an orbital period of 242 days
({\em Mathieu et al.}, 1991) and a separation slightly more than
1~AU. These authors used a circumbinary disk model to reproduce the
mid-infrared excess at $20\,\mu$m: GW~Ori is one of those
spectroscopic binaries in which a large emitting region is needed to
explain the sub-mm flux.  With an estimated stellar separation of
$\sim1$~AU, this requires an extended circumbinary structure.  The
presence of circumbinary material was even confirmed by {\em Mathieu
et al.} (1995) who found that independently of any specific disk
model, the extended ($\approx 500\,$AU) sub-mm emission of GW~Ori was
circumbinary in origin.

\subsubsection{DQ Tau}
Like GW~Ori, this $0.1$~AU separation spectroscopic binary possesses
excess emission at longer wavelengths, indicating the presence of circumbinary
material around the central stars.  Further observations have revealed 
evidence for accretion bursts near the binary periastron in the form of
photometric variability ({\em Mathieu et al.}, 1997) and 
increased veiling ({\em Basri et al.}, 1997). These results are consistent with 
the prediction by {\em Artymowicz and Lubow} (1996), who showed
that accretion streams are likely to link the inner edge
of the circumbinary disk to the stars. Thus DQ~Tau is an example of a binary where
replenishment from a circumbinary structure is at work. 

\subsubsection{V4046 Sgr}
This pair has an  orbital period of 2.4~days and an eccentricity close to zero. 
{\em Artymovicz and Lubow} (1996)'s models of accretion from the
circumbinary environment predict that mass ratio, $q$
(M$_2$/M$_1$), $\sim$1, low-eccentricity
binaries should not experience accretion bursts. However,
{\em Stempels and Gahm} (2004) have recently observed spectroscopic features
that can be explained by the presence of gas concentrations in corotation
with the central binary. These gas accumulations might provide further
evidence for accretion from the circumbinary environment. 

\subsubsection{AK Sco }
AK Sco is an eccentric spectroscopic binary with q$\sim$1 and
a separation of 0.14~AU. The circumbinary disk needed to explain the
spectral energy distribution possesses an inner hole
of radius $\sim$0.4~AU within which the binary resides.
This is consistent with the prediction of {\em Artymowicz and Lubow} (1996) 
for the inner rim of a circumbinary disk in such a system.
Like DQ~Tau, it also shows evidence of accretion bursts related to the
orbital motion, but not near periastron ({\em Alencar et al.}, 2003). Indeed,
the H$\alpha$ equivalent width peaks at the orbital phase when the
stars are farthest apart. \\

These puzzling results show that the search for clear signs of
circumbinary accretion onto the central system of young binaries is
on-going. However, if circumbinary environment replenishment occurs
only when the binaries  are sufficiently
close, imaging such systems will be very difficult. Future interferometric
measurements might allow us to disentangle the various possible modes of accretion.

\section{DISK EVOLUTION IN YOUNG BINARIES}
\label{sec:evol}

\subsection{The Need for Resolved Observations of Young Binaries}

  A problem with using ensembles of T Tauri stars
for discerning evolutionary trends is that one has to
make judgements about the ages of the stars concerned. Some
studies have used pre$-$main-sequence evolutionary tracks
to ascribe ages to individual systems (e.g., {\em Hartmann et al.},
1998; {\em Armitage et al.}, 2003), whereas others simply assumed that
all stars in a given star forming region have a similar age (e.g.,
{\em Haisch et al.}, 2000). In each case, the assignment of age is
subject to uncertainties as a result of both the uncertainties in the
pre$-$main-sequence tracks and the additional errors
introduced by placing unresolved systems, as opposed to
individual stars, in the HR diagram.

  In binaries, however, we know {\it a priori} that the
components are coeval, at least to within $\sim 10^5$
years (i.e. to within a small fraction of the average
ages of T Tauri stars). This statement is based
on theoretical models for binary formation:
the only possibility for binaries forming in a significantly
non-coeval fashion is via star-disk capture. A number of
studies have however shown that this is likely to be a very
minor source of binary systems, even in dense environments like
the Orion Nebula Cluster ({\em Clarke and Pringle}, 1991; {\em Scally
and Clarke}, 2001). Therefore, without any need to rely on the accuracy
of pre$-$main-sequence tracks, we can use binary stars as
stellar pairs that are 
guaranteed to be coeval. 

In recent years, each of the diagnostics  described in section~\ref{sec:diag} has been used
extensively to study the timescale and nature of evolutionary processes
in protostellar disks. Typically these studies have not separated the
individual components  in binaries closer
than an arcsecond or so. Because closer binaries constitute more than
half of the systems in the best studied region, Taurus Aurigae, this means
that conclusions on disk evolution based on these studies are subject
to considerable uncertainties. 

    For example, the designation of spectral
types, and hence masses, to unresolved systems is unreliable; likewise,
the detection of a disk diagnostic in an unresolved system
does not in itself indicate whether  it is the primary or the
secondary or both components that possess a disk. These two factors
introduce considerable uncertainties when using such data to investigate
how disk evolutionary processes depend on stellar mass. 

 Another  potential problem resulting from using
unresolved data  relates  to the case in which
the distribution of some observed property in T Tauri systems
is used to infer the {\it rate} at which systems pass through
various evolutionary stages. Evidently, this analysis is compromised in the
case that the observed property is the sum of quantities arising 
from the individual binary components, whose evolution may not
be synchronized. For example, 
the distribution of T Tauri stars in the K-L, K-N  two colour plane
has been used to deduce the relative amounts of time that stars spend
with disks that are respectively optically thick or   
optically thin (``transition disks'') or undetectable 
({\em Kenyon and Hartmann}, 1995). This study
revealed the striking result that very few systems were located
in the transition region of the two colour plane, and has motivated
the quest for disk clearing models that can effect a 
rapid
dispersal of the inner disk ({\em Armitage et al.}, 1999; {\em Clarke et al.}, 2001;
{\em Alexander et al.}, 2005).  {\em Prato and Simon} (1997) recognised that
interpretation of this diagram is complicated by the existence of
binaries and argued that the small numbers of systems with colours
characteristic of transition objects implies that mixed binary
pairs (i.e. one star with a disk and one without disk) must be relatively rare.
Our analysis in
\S\,\ref{sub:dist-aq} below shows that mixed pairs do in fact occur quite
frequently in systems whose components have very disparate masses;
in this case, however, the infrared colours of the unresolved system
are then dominated by that of the primary and so such
systems do not frequently end up in the transition region.

  In summary, although studies of disk evolution based
on unresolved systems are indeed valuable, they represent a
rather blunt instrument compared with that provided by studies
that resolve the individual components of binary systems. 
The value of this latter data can only be exploited if we first
use it to answer a fundamental question: to what extent is
disk evolution affected if the disk in question is located in a
binary system? Depending on the answer to this question, we can
{\it either} use the data to explore the influence of binarity on disk
evolution {\it or} use the binary environment as just representing
samples of coeval stars of various masses. We will return to
this issue in section \ref{sub:dist-aq} below.

\subsection{Overview of the database: separation distribution
of binaries and associated selection effects }

 We have classified the binaries in the database for which we have
been able to assess the presence of a disk in each component as CC,CW,
WC and WW. Here C denotes a cTT (accreting, disk possessing) star and
W a wTT (non-accreting, generally diskless) star.  The first and
second letter refer to the primary and secondary, respectively. The
designation of C or W for each component is based primarily on the
criterion of {\em Mart\'{i}n} (1998) for the equivalent width of
H$\alpha$ as a function of spectral type.  In the minority of systems
for which this is not available, the presence of Br$\gamma$ is used
instead. In the absence of information on either of these diagnostics
a cut-off in near-infrared color of $K-L = 0.3$ or mid-infrared color
of $K-N = 2.0$ is employed instead.  We also consider two additional
categories, CP and WP, in which the primary is a cTT or a wTT and the
secondary is a ``passive disk'' object; a non-accreting star that
while generally lacking any near-infrared excess also possesses a
significant mid-infrared excess, indicating the presence of an inner
dust disk hole ({\em McCabe et al.}, 2006).

  Table~\ref{tab:cc-dist} lists the numbers of objects of each type in the database
that satisfy certain criteria. The left hand column lists
the number of objects of each type that have the most complete
information (i.e. binary separation and spectral type for each component).
Objects in the left hand column have not been reported as possessing
additional unresolved companions (at $< 0.1"$ separation) to one
of the components, a feature which would disrupt the accretion flow in that
region. The second column (which includes those in the first column)
covers the larger sample of systems with known
separations but not necessarily spectral types for both components.
Objects in this column also have no reported additional close companions. The third
column lists the number of systems with additional close companions.

\begin{table}[htb]
\caption[]{Numbers of binaries in the database according to classification; see
text for details.\label{tab:cc-dist}}
\begin{flushleft}
\begin{center}
\begin{tabular}{cccc}
\hline\hline
CC & 29  & 38 &7\\
CW & 11 & 14  &1 \\
CP & 2 & 2 & 0\\
WC& 4 &6  &1\\
WP& 1 & 1  &0\\
WW& 12 & 21&1 \\
\hline
\end{tabular}
\end{center}
\end{flushleft}
\end{table}

  To some extent, the numbers in Table~\ref{tab:cc-dist} reflect observational
selection effects. For example, it is possible that binaries
with W primaries are under-represented in this sample: comparison 
of in Table~\ref{tab:cc-dist} with the total numbers of stars
in the Taurus aggregates that are classified as cTTs and wTTs,
100 and 70, respectively (Guieu, private communication),
suggests a mild deficit of binaries with wTT primaries.
Any under-representation is likely to result from
the relative disincentive to make  high angular resolution observations
of objects which show no obvious accretion signatures in their
combined spectra. We would expect this under-representation
to be more acute at small separations (where resolved
observations require more effort) and, in the case of WCs,
in low-mass ratio objects, where the accretion signatures of
the secondary are not obvious in the combined spectrum. In
addition, relatively few objects have been scrutinized
at $N$-band, so that further systems may subsequently
be transferred from the CW/WW to the CP/WP category; we have
been rather conservative in our assignment of passive systems
in Table~\ref{tab:cc-dist}, and so have not included several
systems judged to be marginal passive candidates according
to {\em McCabe et al.} (2006).



\begin{figure}[htb]
 \epsscale{1.0}
\plotfiddle{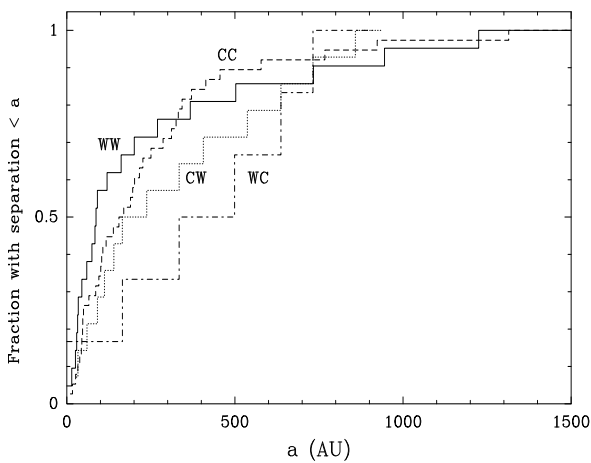}{0pt}{-90}{170}{220}{0}{0}
 \caption{Cumulative separation distribution of the four different binary category.\label{fig:cumdist}}  
 \end{figure}

 In Fig.~\ref{fig:cumdist} we plot the cumulative separation distributions
of the binaries in the central column of Table~\ref{tab:cc-dist}, with the 
histograms (in descending order at $a=500$~AU) representing
CCs, WWs, CWs and WCs. There is no statistically significant
difference between any of these distributions: in the case of
the two categories of binary with the largest sub-sample
numbers, the CCs and the WWs, a KS test indicates that in the
case that the two sub-samples were drawn from the same
parent distribution, the
probability that the samples would be at least as different
from each other as observed is $25 \%$ . There is some
theoretical expectation that disk evolution should be accelerated in
closer systems (see below), which might in principle lead
to an excess of WWs at small separations. Although the fraction
of close binaries is somewhat higher for WWs (i.e. $57 \%$ of
WWs have separation less than $100$~AU compared with
only $38 \%$ of CCs), this difference is not 
statistically significant, possibly implying
that accelerated disk evolution at small separations is not 
occurring in the binaries in our sample, which
are rarely closer than $\sim 20$~AU. On the other hand, as we mentioned
above, there is an  observational selection effect against the 
discovery of closer systems with a W primary, so that this
might mask any evidence for accelerated disk evolution in
closer binaries.  

  Fig.~\ref{fig:cumdist} also demonstrates that mixed pairs (WCs, and, to a lesser
extent, CWs) are more concentrated at larger separations, although
again the relatively small numbers of these systems yields a
statistically insignificant result.  The KS probability of
either the mixed binary samples having a different
separation distribution from the CC or WW samples is never 
less than $25 \%$. We are less inclined
to ascribe this tendency to an observational selection effect,
since there is no reason why WCs should be under-represented
at small radii compared with WWs, or why CWs should be under-represented
compared with CCs at small separations.

  The numbers of mixed systems (CWs or WCs) compared with CCs is a measure
of the difference in lifetimes of the disks around each component.
Synchronized evolution would imply mixed systems should be very rare,
whereas a large difference in lifetimes would imply that mixed
systems should be abundant. Including also the 4 passive systems
as mixed systems, the total numbers of CCs compared with
mixed systems is 37 compared with 24; we have avoided
the complicating factor of close companions by using the systems
in the middle column of Table~\ref{tab:cc-dist}. This implies that the average
lifetime of the shorter lived disk is $\sim60 \%$ of the longer
lived disk. A further point to make about the mixed systems
is that the number of mixed systems with a cTT primary
compared with a wTT primary is 17 compared with 7.  Evidently,
there is a tendency for the primary's disk to be longer lived,
although this is not universally the case.

  We therefore conclude that when one combines all the available
data from the literature, mixed systems are much less rare
than was previously thought. It would appear that the reason
that we need to revise our conclusions is that the incidence of
mixed systems varies between different star forming regions
(see also {\em Prato and Monin}, 2001).
Thus among the CCs and WWs in the middle column of Table~1,
around half are located in Taurus. However, only $20 \%$ of the
mixed systems are located in Taurus. Thus early studies (e.g., {\em Prato
and Simon}, 1997) whose targets were mainly in Taurus contained
relatively few mixed systems. We can only speculate as to why the
fraction of mixed systems should vary from region to region. One
obvious possibility is if the mixed phase corresponds to a particular
range of ages and if different star forming regions have different
fractions of stars in the relevant age range.

\subsection{The distribution of binaries in the $a-q$ plane}
\label{sub:dist-aq}

\begin{figure}[htb]
 \epsscale{1.0}
\plotfiddle{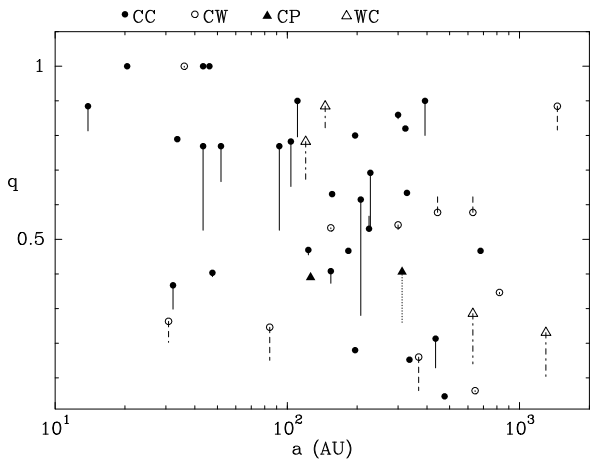}{0pt}{-90}{170}{220}{0}{0}
 \caption{Binaries from the left hand column of
Table 1 plotted in the $q,a$ plane \label{fig:dist-cc}}  
 \end{figure}

 To make further progress, we must examine how various categories
of binaries are distributed in the plane of mass ratio versus
separation. This necessitates using the more restricted sub-sample
listed in the left hand column of Table~\ref{tab:cc-dist}, for which we have
spectral type information for each component. We have checked
that the separation distribution of the sub-sample is consistent
with that of the full sample; although the difference is not
statistically significant, we note that there happens to be
a deficit of wide ($> 500$~AU) CC binaries  in the sub-sample
compared with the full sample, which is manifest as the lack
of solid dots in the right hand portion of Fig.~\ref{fig:dist-cc}. 

 In Fig.~\ref{fig:dist-cc},  filled circles  represent the CCs, open
circles the CWs, filled triangles
the CPs and open triangles the WCs. We do not include the WWs in this plot
since they contain no information about {\it differential} disk
evolution. We note that we expect the selection effects to be similar
for all the binaries with cTT primaries and that we expect the selection 
bias against low $q$
and low $a$ systems to be more severe for the systems with wTT primaries.   

 We have placed binaries in Fig.~\ref{fig:dist-cc} using the
correlation between spectral type and mass for stars of age
1~Myr given in {\em Hillenbrand and White} (2004).  The necessity of
having an optical spectral type for each star means our
sample of  
29 CCs and 11 CWs has excluded any binary
containing an infrared companion or Class I source. For each binary we
then calculate $q_{\rm DM}$ (i.e. the mass ratio $M_2/M_1$)
using the  pre$-$main-sequence tracks of {\em D'Antona and Mazitelli} (1994).
For a subset of systems for which both spectral types are later than
K3, we also compute $q_{\rm BCAH}$, using the pre$-$main-sequence tracks
of Baraffe et al. (1998), also listed in {\em Hillenbrand and White} (2000).
In Fig.~\ref{fig:dist-cc}, we plot 
$q_{\rm DM}$ in each case
but link $q_{\rm DM}$ to the corresponding value of $q_{\rm BCAH}$
in the systems where both components lie in the range where
$q_{\rm BCAH}$ can be computed. We use different dashes for different type of pairs.
The length of the vertical lines
gives some indication of the uncertainties inherent in
pre$-$main-sequence tracks, although cannot in any sense be
regarded as an errorbar on $q$. Despite the strong
disagreement between the tracks in certain ranges of spectral
type, we nevertheless find that both set of tracks are
in broad agreement as to whether binary systems are high
or low $q$. In the quantitative analysis of the $q$ distributions
described below, we use $q_{\rm DM}$ as this is the only
quantity that is available for all systems in our sample.

  There are several striking features in this figure. 
As we have already noted, it first demonstrates that
mixed systems 
are not rare and that many of
the mixed systems are binaries with low $q$. On theoretical
grounds (see below), one might expect that systems where the secondary's
disk is exhausted before the primary's (i.e. the CWs and the
CPs) would be low $q$ binaries. This is borne out with marginal
statistical significance when one compares the $q$ distribution
of the CCs with the combined population of CWs and CPs.  If
we restrict our sample to binaries closer than $1000$~AU in
order to reduce the risk of picking up chance projections in our
sample, we find that a KS test reveals that the two $q$ distributions
are different at the $2~\sigma$ level.  A KS test assesses
the statistical significance of the {\it maximum} difference
between the two datasets, which in this case refers to the
fact that $11/28$ CCs have $q < 0.6$ whereas for CWs and CPs
the combined figure is $11/13$.  We also note that systems
in which the primary's disk is exhausted first are relatively
rare, i.e. for $a < 1000$~AU the total number of WCs and WPs
is 4, compared with the 13 mixed systems with a cTT
primary in this separation range. From Fig.~\ref{fig:dist-cc}, we see that
these 4 mixed systems with wTT primaries are 
not found
preferentially at low $q$, in contrast to what appears
to be the case for the mixed systems with cTT primaries. 
However, we caution that there may be a selection effect against
the detection of low $q$ mixed systems with wTT primaries at small
separations.  

  Further analysis of this figure (i.e. division of the $(a,q)$
domain into different regimes) is rendered difficult by the
small total number of objects, so any trends that might appear
to be qualitatively significant do not correspond to an impressively 
significant KS statistic. For example, we draw attention to the
fact that for binaries closer than $100$~AU, this being
the canonical scale of disks around young stars
({\em Vicente and Alves}, 2005; {\em McCaughrean and Rodmann}, 2005), there are
{\it no} examples of pairs in which the primary's disk is
exhausted first (i.e. WCs or WPs) and that $2/3$ of the mixed
systems have $q < 0.5$ compared with only $2/10$ of the CCs
having such low values of $q$. 

\begin{figure}[htb]
 \epsscale{1.0}
  \plotone{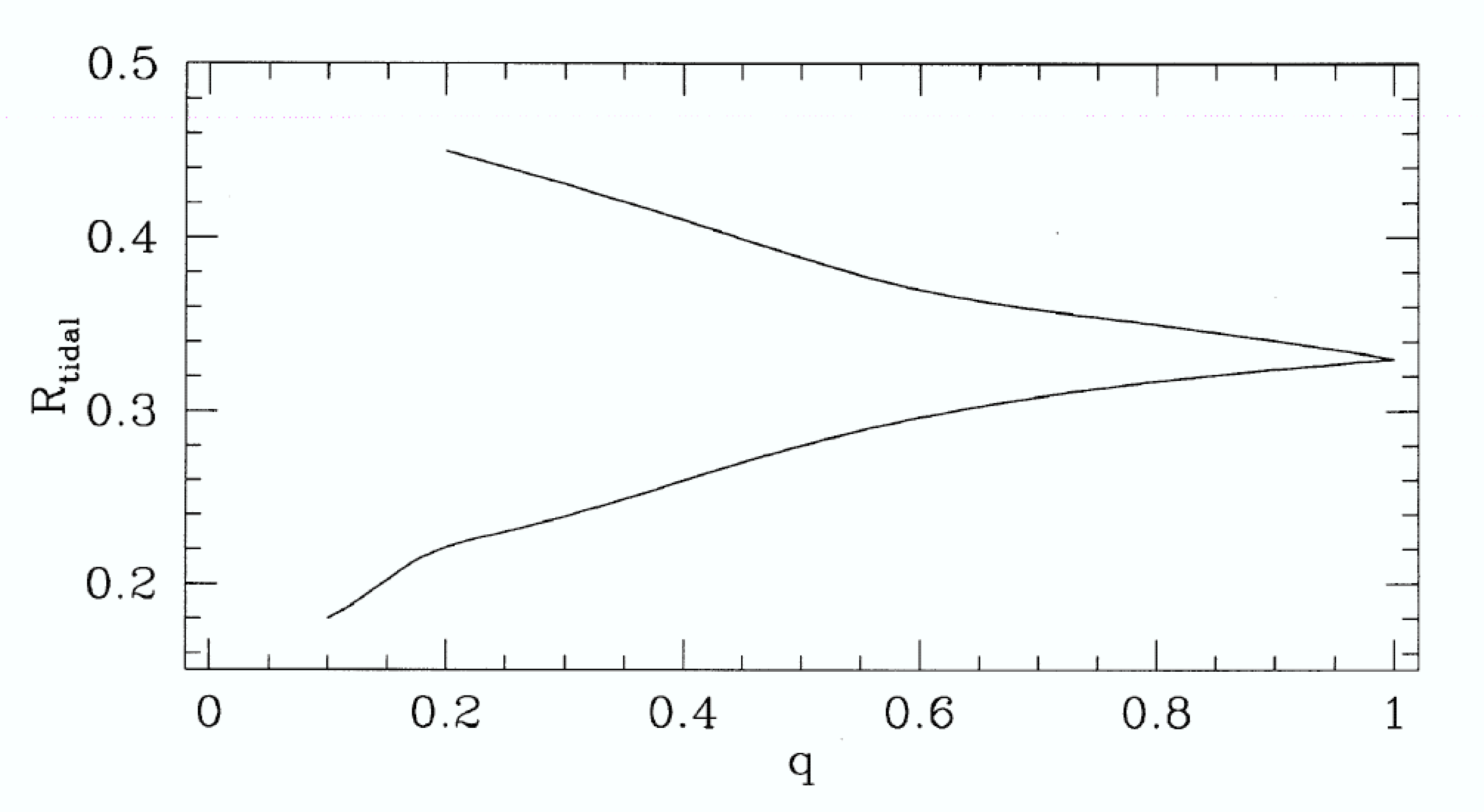}
   \caption{Truncation radius as a fraction  of the semimajor axis of the binary
 orbit vs $q$: upper line for primary, lower line for
secondary (from {\em Armitage et al.} 1999).
   \label{fig:fqrad}}
 \end{figure}

 This behaviour is qualitatively
consistent with what is expected theoretically in the
case in which the disks around each star evolve in isolation,
with their outer radii set by tidal truncation
in  the binary potential. 
Tidal truncation of disks occurs at a radius equal to
a factor $R_{\rm tidal}$ times the binary separation, where $R_{\rm tidal}$
is plotted in Fig.~\ref{fig:fqrad}
({\em Armitage et al.}, 1999; {\em Papaloizou and Pringle}, 1987).

Evidently, for binaries at fixed
separation, the secondary's disk is always tidally
truncated to a smaller radius, but the difference only
becomes significant for $q$ less than about $0.5$.
In the case of disks that are not continually
replenished from an external reservoir, the tidal
limitation of the disks around secondaries at a smaller
radius leads to a more 
rapid accretion of
the secondary's disk ({\em Armitage et al.}, 1999).
This can be readily understood, as disk accretion
depends on viscous redistribution of angular momentum, which,
in a freely expanding disk, occurs on a longer and longer
timescale as the disk spreads outwards. If a disk is tidally
truncated, however, angular momentum is tidally transferred
to the binary orbit at the point that the disk grows to
the tidal truncation radius. Hence the disk dispersal
timescale is roughly given by the disk's viscous timescale
at the tidal truncation radius. For a disk with surface density
profile of the form $R^{-a}$, the viscous timescale at
radius $R$ scales
roughly as  $R^{2-a}$.  Hence, for $a$ in the range $1-1.5$
({\em Beckwith and Sargent}, 1991; {\em Hartmann et al.}, 1998), we have that the
viscous timescale at the tidal radius $R_T$ scales
as $R_T^{0.5-1}$. Putting this scaling together with
Fig.~\ref{fig:dist-cc}, we can therefore see that for binaries with
$q > 0.5$, the viscous timescales at $R_T$ are sufficiently
similar that the disks should evolve more or less synchronously.
The phase during which the secondary has exhausted its disk,
but the primary has not, is relatively brief. On the other hand,
for lower $q$s in the range observed, we expect the viscous
timescales at $R_T$ for the two components to differ by
order unity. This means that the time spent by a system
as a CW is comparable with the time spent as a CC, and hence,
as observed, the two sorts of system should occur in roughly
equal numbers.

  At larger separations, $ a > 100$~AU, the picture
is apparently rather different since now mixed systems
with wTT primaries start to appear. This suggests that
we are now entering a regime where the tidal truncation
condition exerted by the binary is no longer the critical
factor in determining which disk is exhausted first, a 
result that is perfectly comprehensible in the limit
that the binary separation is much larger than typical
disk sizes. We also note that the data  for the wider
binaries (where the disks evolve without obvious reference
to their location in a binary) provides
good evidence that 
disk lifetime is not a strong function of  stellar mass. As an example,
Sz 30 and Sz 108 are mixed systems with identical separations
($630$~AU) and similar spectral types for each component
(M0.5-M2 and M0-M4.5 respectively). Nevertheless, in the former
system it is the secondary that has lost its disk and in the latter
it is the primary.   
Because we cannot appeal to non-coevality
to explain this difference, we must assume that the
lifetime of isolated disks is not a strong function of
stellar mass in the range $0.1-1 M_\odot$, and, hence,
that presumably the initial conditions in the disk (such as
initial mass or radius) instead dictate disk lifetime.

\subsection{Implications for disk resupply}

 Early studies of binaries in which accretion diagnostics  were
separated for each component concluded that mixed systems are
rare (see discussion in {\em Prato and Monin},  2001), leading
{\em Prato and Simon} (1997) to argue
that the disks around each component must be sustained and then dissipated 
in a synchronised manner. It is hard to understand synchronised dispersal
unless it is effected by some external agent. 
On the other hand, a low fraction of mixed systems can be explained if
both components are fed from a common reservoir over most of the
disk lifetime and if, once the reservoir is exhausted, the dispersal
of both disks is relatively rapid. This explanation was favoured
by {\em Prato and Simon} on the grounds that continued replenishment
is the only way to explain the presence of accretion diagnostics 
in the closest binaries ($a <$ a few~AU), for which the viscous
timescale of their (highly truncated) disks is much less than the
system age. In these closest binaries, there is good evidence
for circumbinary disks ({\em Jensen and Mathieu}, 1997), 
which can plausibly continue to feed the central binary ({\em Mathieu
et al.}, 1997). In wider binaries, however, i.e. $a$ in the range a few to
$\sim 100$~AU, upper limits on circumbinary disk masses
are $\sim 5$ Jupiter masses ({\em Jensen et al.}, 1996) and therefore inadequate
to provide substantial replenishment of circumstellar disks. In these
wider systems, it is instead  necessary to invoke replenishment
through infall from an extended
envelope. Possible evidence for such an envelope is provided by the
millimeter study of young binaries by {\em Jensen and
Akeson} (2003) who found that their interferometric measurements 
contained $46-85 \%$ of the flux found in previous, single dish 
measurements ({\em Beckwith et al.}, 1990).  {\em Jensen and Akeson} therefore speculated that
the additional flux originated in an envelope on scales of $>700$~AU, with the caution
that the flux difference  could be due to 
a flux calibration issue.  However, as it is possible to conceal large quantities
of cold dust at large distances from the binary without contributing significantly
to the millimetre flux ({\em Lay et al.}, 1994), it is impossible to use this
observation to constrain whether the extended emission contains
a viable mass for re-supplying the binaries' circumstellar disks.

  Our analysis here however indicates that mixed systems are, in fact,
common, and thus does not require continued replenishment
of disks for the binaries in our sample (which mostly have
separations $> 20$~AU). Our results do not require there
to be no replenishment, but imply that such replenishment must occur
over a minor fraction of the disks' lifetimes or else be concentrated
on to the primary's disk at late times. This latter possibility
is in conflict with numerical simulations of infall onto proto-binaries
({\em Artymowicz}, 1983; {\em Bate}, 1997; although see {\em Ochi et al.}, 2005 for a recent contrary
view on this issue). The simplest interpretation of our results, however,
is that the disks evolve in isolation and that disk tidal truncation
in the binary potential results in the secondary disk being dissipated
somewhat prior to the primary's disk.

\subsection{Implications for planet formation
in binaries}
\label{sub:planets}

  How do these findings bear on the probability
that planets are located in binary systems? The presence of a
binary companion may render the existence of planets less likely
in two ways. First, binarity restricts the regions of
orbital parameter space in which planets can exist in stable,
circumstellar orbits, ruling
out orbital radii that are within a factor of the binary separation,
modulo the mass ratio $q$.
For example, {\em Holman and Weigert} (1999) have conducted a 
study of  the long term orbital stability of planets in binary systems
and find that a companion star orbiting beyond more than 5 times 
the planetary orbital radius 
does not strongly threaten the planet's orbital stability. Second,
if binarity reduces disk lifetimes (in the
primary or secondary or both) then it may reduce the probability
of planet formation, since there may be insufficient time
for slow processes (such as those involved in the core accretion model)
to operate before the disk is dispersed. For example, 
{\em Thebault et al.} (2004) find that the formation of the observed
planet at $2$~AU 
in the $18$~AU binary $\gamma$ Cephei requires the presence of a long lived
and massive gas disk.
In the absence of such gas, 
secular perturbations
by the binary companion generate too high a velocity dispersion
among the planetesimals for runaway accretion to proceed.

  The present study, however, finds that the influence of binarity
on circumstellar disk lifetime is rather mild in the systems
with separations $> 20$~AU. The fact that the separation
distribution of diskless binaries is indistinguishable from that
of binaries with disks suggests that disk dispersal is not
strongly accelerated for the closer binaries in this sample.
Concerning differential evolution between the disks around primaries
and secondaries, we found that the overall statistics 
of mixed systems versus CC systems implied that the
shorter lived disk (usually the secondary's) had a mean lifetime
of $\sim60 \%$ that of the longer lived disk. Unless there are
processes in planet formation for which a factor $2$ difference
in disk lifetime is critical, we conclude that {\it circumstellar
planet  formation 
is not likely to be strongly suppressed in the case of binary secondaries}.
We therefore expect planets to be formed around
both components in binary systems wider than $\sim 20$~AU. 
The recent numerical simulations of {\em Lissauer et al.} (2004) and
{\em Quintana et al.} (2005) (see also
{\em Barbieri et al.}, 2002) are in good agreement with this result.

  The observational situation regarding the detection of planets in
binary systems is strongly skewed by the selection criteria used in 
Doppler reflex motion surveys,
as these tend to exclude known binaries on the grounds
that binary orbital motion makes it harder to detect
a planetary companion. Among the more than a hundred and fifty 
G to M stars hosting planetary companions, only 25 are binary or multiple
systems, hosting a total of 31 planets (exoplanets.org; 
{\em Eggenberger  et al.}, 2004, 2005; {\em Mugrauer et al.}, 2005).
Therefore, only around $15 \%$ of
known planets are in binary or multiple systems.   

\begin{figure}[htb]
 \epsscale{1.0}
  \plotone{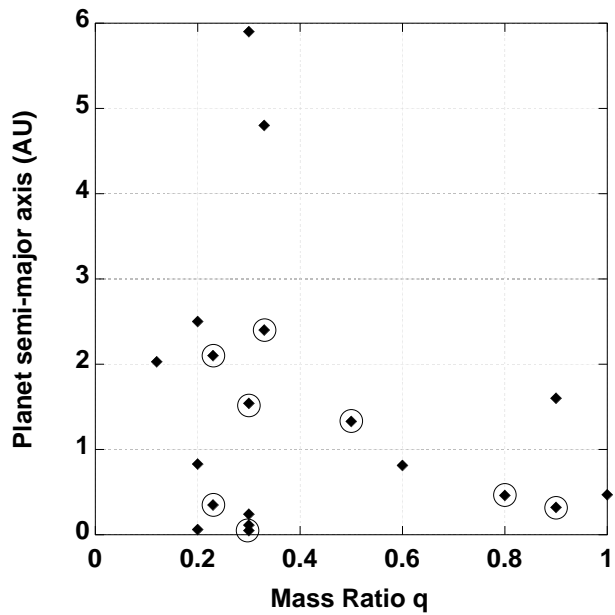}
   \caption{Distance from the planet to its central star (component of a binary) 
     vs mass ratio. The encircled points are the ones for binaries with separation less than 500~AU.
     \label{fig:dist-vs-q}}  
 \end{figure}

\begin{figure}[htb]
 \epsscale{1.0}
  \plotone{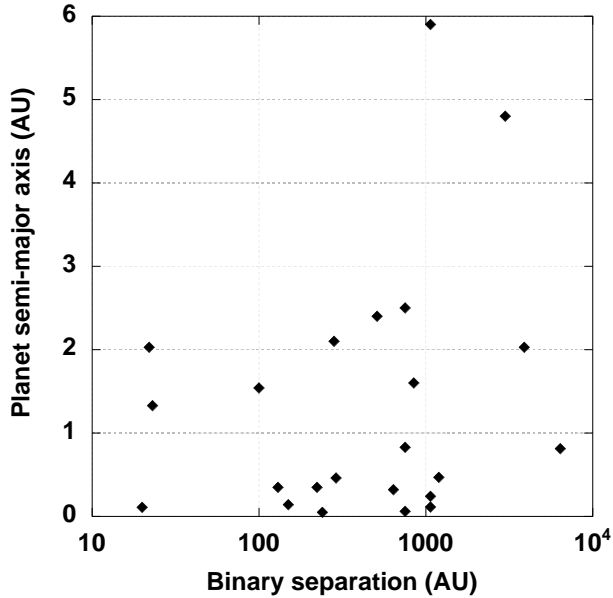}
   \caption{Distance from the planet to its central star (component of a binary) vs binary separation
 \label{fig:dist-vs-sep}}  
 \end{figure}

Fig. ~\ref{fig:dist-vs-q} 
and ~\ref{fig:dist-vs-sep}
show the orbital properties of the binary systems known to
host Doppler reflex motion planets. As expected, the sample is strongly
biased towards larger separations: planet search programs
do not typically monitor binaries with separations less than
$\sim2''$, corresponding to separations in the range $>2$~AU
at the distances of the target stars ({\em Valenti and Fischer}, 2005).
Because the  median binary
separation for G stars is $30$~AU ({\em Duquennoy and Mayor}, 1991),
it is evident that a large fraction of binaries have been excluded
from such surveys. There is also the possibility of an observational bias
towards low
$q$ on the grounds that low-mass companions are more likely to have
been overlooked when initially selecting the radial velocity
targets. 

  It is immediately obvious from Fig.~\ref{fig:dist-vs-sep} that the ratio of binary
semi-major axis to planet semi-major axis ($a_b/a_p$) is extremely large,
generally in the range $100-1000$ and in all cases $>10$. It is
therefore unsurprising, on the grounds of orbital stability,
that planets are found in these systems. Moreover, the binaries
in Fig.~\ref{fig:dist-vs-sep} are in the same separation range that we have studied
in Section 4.2, where we found little apparent dependence of
disk lifetime on binary separation. We would therefore not
expect planet formation to be suppressed in these systems
on the basis of reduced disk lifetime.

We stress that  the current data
cannot be used to determine whether planets
are preferentially found around binary primaries or secondaries,
since in almost all cases it is only the primary that has been
a radial velocity target. In only two systems is the planet detected
around the secondary component (16~Cyg and HD~178911). 
Likewise, it would be  premature to derive
the statistics  of {\it circumbinary planets}. To date, there
is one system, HD 202206, that might be described as containing
a circumbinary planet, although the mass ratio of the central
binary is extremely low: the central companion is itself in the
brown dwarf/planetary regime ({\em Correia et al.}, 2005).
From a theoretical point of view, {\em Moriwaki and Nakagawa} (2004)
have claimed that in the case of a binary of separation 
$1$~AU, planetesimal accretion should be able to proceed undisturbed
at radii greater than $\sim 13 $~AU from the barycentre. This
relatively large region in which planet formation might be expected to
be suppressed in the circumbinary disk means that it may be problematic
to detect planets through radial velocity measurements around all but
the closest binaries.  {\em Quintana et al.} (2005) calculate, however, that
for binary separations of $<$0.2~AU, the growth of planetesimals into a
system of terrestrial planets is statistically indistinguishable from similar
simulations for single stars.
Surveys for planets around single-lined, spectroscopic binaries
(e.g., {\em Eggenberger et al.}, 2005) have only recently begun. 
When data are available, they should provide interesting constraints.

\section{FUTURE DIRECTIONS}
\label{sec:future}

The most formidable obstacle to furthering our
understanding of disk 
evolution in young binaries is
the relatively small size of our database. Although
our compilation of around 60 binaries with complete
spectral type and disk diagnostic information for
each component   represents
tremendous progress in the last decade, it is nevertheless
too small a sample for us to be able to divide it
into sub-categories according to, for example, separation
and subsequently derive statistically significant results.
There are however good prospects for increasing the
sample size.  In our database
of $\sim$170 total systems, we estimate that we can derive 
complete properties for approximately another half dozen
systems based on extant data.  An additional 28 systems 
with separations of $>$1$''$ can
be characterized with a 2$-$3~m class telescope in a site
with good seeing, such as Mauna Kea or Cerro Tololo.
A further 3 dozen systems have separations between 0.1$''$
and 1.0$''$.  For these pairs it would be straightforward to
characterize each component with low-resolution spectroscopy behind an
adaptive optics system, or an integral field spectrometer
unit, at a 6$-$10~m class facility.  The results of such
observations would more than double the young binary sample.
Furthermore, our database was compiled from a limited number
of references and is certainly far from complete.  We anticipate
the on-going compilation of additional objects and improvement
in the quantity and quality of data for objects already listed.

Larger samples of binaries with known properties in 
a variety of star forming regions with a range of estimated
ages will allow us to test the extremely intriguing notion
of the regional dependence of the fraction of mixed systems.
The data in this paper, as well as data obtained in the earlier studies of
{\em Prato and Simon} (1997), {\em Prato and Monin} (2001), and
{\em Hartigan and Kenyon} (2003), suggest a low fraction of
mixed pairs in the Taurus region.  Could this be the result
of a younger age for Taurus than the other regions from
which our sample is culled?  Is it simply a selection effect,
or a result of small number statistics?  If a real and
age-dependent effect, the mixed system fraction may 
yield a unique and sensitive approach to estimating the ages of
star forming regions.

With high-resolution spectroscopy of both components in young
binaries more detailed properties may be examined.  For example,
with multiple epoch observations hierarchical spectroscopic binaries
might be identified in binary component stars.
The individual rotation properties of the stars in close pairs
could also be examined and compared with the circumstellar disk
properties to better understand the
evolution of angular momentum in young binaries ({\em Armitage et al.}, 1999).
High-resolution observations of accretion line diagnostics, 
such as hydrogen emission lines, could provide a unique
approach to the measurement of how accretion is apportioned 
between the two stars in spectroscopic binaries.  Such 
observations at infrared wavelengths would provide a better
opportunity to observe emission lines from both stars, even
for systems with large continuum flux ratios (e.g., {\em Prato et al.}, 2002).

An interesting problem raised in {\em McCabe et al.} (2006) is the
origin of the passive disk phenomenon.  By combining 
resolved near- and mid-infrared observations with longer
wavelength Spitzer data and astrophysical information for
the binary stars themselves, i.e. masses, it will be
possible to test the premise set up in {\em Clarke et al.} (2001)
and {\em Takeuchi et al.} (2005), namely that a population of
young systems with large inner disk holes exists around
higher mass stars that have previously been identified as wTTs.

The advent of very high resolution interferometry, in both the
optical-infrared as well as in the millimeter regimes, will
provide an unprecedented view of the orientations of disks
in binaries even at circumstellar scales.  Already progress
has been made using the Keck Interferometer ({\em Patience et al.}, 2005)
and the VLTI ({\em Malbet et al.}, 2005).
The ALMA interferometer, anticipated for first light in the 
next 3$-$4 years at partial capacity, will provide
unprecedented images of the cool, dusty disk structures.

These new generations of facilities will enable entirely new
studies, which will go far beyond the issue of simple existence
of disks in binary systems. Instead it will be possible
to measure how {\it disk properties} vary as
a function of binary properties such as
separation, mass ratio, angular momentum, magnetic field
strength, etc.
For example, an instrument such as ALMA will
enable us to study disk particle size distributions as a
function of binary separation.  Optical-infrared interferometers
could provide data on inner disk structure as a function of
magnetic field strength.  Numerous such exciting
possibilities for future study exist.

 \bigskip

\textbf{ Acknowledgments.} We thank the anonymous referee for useful comments that helped
improve the quality of the paper. C.~C. thanks the LAOG and J.-L.~M. thanks the
IoA for their generous hospitality while writing this paper.  L.~P. acknowledges the
contribution of helpful information from K. Strom, S. Strom, and G. Marcy.
We are grateful to A. Eggenberger for discussions of her research in advance of
publication. This work has made use of the ADS database.

\bigskip

\textbf{REFERENCES}
\bigskip
\parskip=0pt
{\small
\baselineskip=11pt
\refs Alencar~S.~H.~P., Melo~C.~H.~F., Dullemond~C.~P., Andersen~J., Batalha~C., Vaz~L.~P.~R. and Mathieu~R.~D. (2003) {\em Astron. Astrophys.,  409}, 1037-1053.
\refs Alexander~R.~D., Clarke~C.~J. and Pringle~J.~E. (2005), {\em Mon. Not. R. Astron. Soc.,  358}, 283-290.
\refs Andrews S.  and Williams J. (2005) {\em Astrophys. J. 619}, L175-178.
\refs Armitage P.~J., Clarke~C.~J. and Tout C.~A. (1999), {\em Mon. Not. R. Astron. Soc.,  304},425-433.
\refs Armitage P.~J., Clarke~C.~J. and Palla F. (2003) {\em Mon. Not. R. Astron. Soc.,  342},1139-1146.
\refs Artymowicz P. (1983)  {\em Acta. Astron., 33}, 223-230.
\refs Artymowicz P. and Lubow  S.~H. (1996) {\em Astrophys. J., 467}, L77-80.
\refs Baraffe I., Chabrier G., Allard F. and Hauschildt P. (1998), {\em Astron. Astrophys.,  337},403-412.
\refs Barbieri  M., Marzar  F., Scholl H. (2002)  {\em Astron.
Astrophys.  396}, 219-224.
\refs Basri G. and Bertout C. (1993) In {\em Protostars and Planets III} (E. H. Levy and J. I. Lunine, eds), pp. 543-566. Univ. of Arizona, Tucson. 
\refs Basri G., Johns-Krull C.~M. and Mathieu R.~D. (1997) {\em Astron. J.,  114}, 781-792.
\refs Bastien P. and M\'enard F. (1988) {\em Astrophys. J., 326}, 334-338.
\refs Bastien P. and M\'enard F. (1990) {\em Astrophys. J., 364}, 232-241.
\refs Bate M.~R. (1997) {\em Mon. Not. R. Astron. Soc.,  285},16-32.
\refs Bate M.~ R., Bonnell I.~ A., Clarke~C.~ J., Lubow S.~ H., Ogilvie G.~ I., Pringle J.~ E. and Tout C.~ A. (2000) {\em Mon. Not. R. Astron. Soc.,  317}, 773-781.
\refs Beckwith S.~V.~W. and Sargent A.~I. (1991) {\em Astrophys. J., 381}, 250-258.
\refs Beckwith S.~V.~W., Zuckerman B., Skrutskie M.~ F., and Dyck H.~ M. (1984) {\em Astrophys. J., 287}, 793-800.
\refs Beckwith S.~V.~W., Sargent A.~I., Chini R.~S.  and G\"ustem R. (1990) {\em Astron. J.,  99}, 924-945.
\refs Bertout C., Basri G. and Bouvier J. (1988) {\em Astrophys. J., 330}, 350-373.
\refs Beust H. and Dutrey A. (2005) {\em Astron. Astrophys., 439}, 585-594.
\refs Bonnell I.~A. and  Bastien P. (1992) {\em Astrophys. J., 401}, 654-666.
\refs Bonnell I.~A., Arcoragi J.-P., Martel H. and Bastien P. (1992) {\em Astrophys. J., 400}, 579-594.
\refs Bohm K.-H. and Solf J. (1994) {\em Astrophys. J., 430}, 277-290.
\refs Brandner W. and Zinnecker H. (1997) {\em Astron. Astrophys., 321}, 220-228.
\refs Breger M. and Dyck H.~M. (1972) {\em Astrophys. J., 175}, 127-134.
\refs Chandler C.~J., Brogan C.~L., Shirley Y.~L. and  Loinard L. (2005) {\em Astrophys. J., 632}, 371-396.
\refs Chelli A., Cruz-Gonzalez I. and Reipurth B. (1995) {\em Astron. Astrophys. Supp., 114}, 135-142.
\refs Chiang E. I.~, Joung M.~K., Creech-Eakman M.~ J., Qi C.,
Kessler J.~ E., Blake G.~ A. and van Dishoeck E.~ F. (2001) {\em Astrophys. J., 547}, 1077-1089. 
\refs Clarke~C.~J. and Pringle J.~E. (1991) {\em Mon. Not. R. Astron. Soc.,  249},584-587.
\refs Clarke~C.~J., Gendrin A. and Sotomayor M. (2001) {\em Mon. Not. R. Astron. Soc.,  328}, 485-491.
\refs Close L.~M., Dutrey A., Roddier F., Guilloteau S., Roddier C., Northcott M., Menard F., Duvert G., Graves J.~ E. and Potter D. (1998) {\em Astrophys. J., 499}, 883-888.
\refs Davis C.~J., Mundt R., Eisloeffel J. (1994) {\em Astrophys. J., 437}, L55-58.
\refs D'Alessio P. (2003) {\em Rev. Mex. Astron. Astrophys.,}, 18, 14-28.
\refs D'Antona D. and Mazzitelli I. (1994) {\em Astrophys. J. Supp., 90}, 467-500.
\refs Donar A., Jensen E.~ L.~ N. and Mathieu R.~D. (1999) {\em BAAS 195}, 1490-1501.
\refs Duch\^ene G., Monin J.-L., Bouvier J. and M\'enard F. (1999) {\em Astron. Astrophys., 351}, 954-962.
\refs Duch\^ene G., McCabe C., Ghez A.~M. and Macintosh B.~ A. (2004) {\em Astrophys. J., 606}, 969-982.
\refs Dutrey A., Guilloteau S. and Simon M. (1994) {\em Astron. Astrophys., 286}, 149-159.
\refs Dutrey A., Guilloteau S., Duvert G., Prato L., Simon M., Schuster K. and M\'enard F. (1996) {\em Astron. Astrophys., 409}, 493-504.
\refs Correia A.~C.~M., Udry S., Mayor M., Laskar J., Naef D., Pepe F., Queloz D. and Santos N.~C. (2005) {\em Astron. Astrophys., 440}, 751-758.
\refs Edwards S., Cabrit S., Strom S.~E., Heyer I., Strom K.~M. and Anderson E. (1987) {\em Astrophys. J., 321}, 473-495.
\refs Edwards S., Ray T., Mundt R. (1993) In {\em Protostars and Planets III} (E. H. Levy and J. I. Lunine eds.) pp. 567-603. Univ. of Arizona, Tucson.
\refs Eggenberger A., Udry S., and  Mayor M.  (2004) {\em Astron. Astrophys., 417}, 353-360.
\refs Eggenberger  et al. (2005) ESO workshop 12-15 july 2005, Garching. 
\refs Eisl\"{o}ffel J. and Mundt R. (1997) {\em Astron. J.,  114}, 280-287.
\refs Eisl\"{o}ffel J., Smith M. D., Davis C. J. and  Ray T.~P. (1996) {\em Astron. J.,  112}, 2086-2095.
\refs Geoffray H. and Monin J.-L. (2001) {\em Astron. Astrophys., 369}, 239-248.
\refs Ghez A. M., Neugebauer G., Gorham P. W., Haniff C. A.,
Kulkarni S. R., Matthews K., Koresko, C. and Beckwith S. V. W. 
(1991)  {\em Astron. J.,  102}, 2066-2072.
\refs Ghez A. M., Neugebauer G. and Matthews K. (1993) {\em Astron. J., 106}, 2005-2023.
\refs Ghez A. M., Emerson J. P., Graham J. R., Meixner M. and Skinner C. J. (1994) {\em Astrophys. J., 434}, 707-712.
\refs Grasdalen G. L., Strom S. E., Strom K. M., Capps R. W., Thompson D., and Castelaz M.
(1984) {\em Astrophys. J., 283}, L57-60.
\refs Gredel R. and Reipurth B. (1993) {\em Astrophys. J., 407}, L29-32. 
\refs Guilloteau S., Dutrey A. and Simon M.  (1999) {\em Astron. Astrophys., 348}, 570-578.
\refs Gullbring  E., Hartmann L., Briceno, C. and Calvet N. (1998) {\em Astrophys. J., 492}, 323-341.
\refs Haisch K. E., Lada E. A. and Lada C. J. (2000) {\em Astron. J.,  120}, 1396-1409.
\refs Haisch K. E., Lada E. A. and Lada C. J. (2001) {\em Astrophys. J., 553}, L153-156.
\refs Hale A. (1994) {\em Astron. J.,  107}, 306-332.
\refs Hartigan P. and Kenyon S. J. (2003) {\em Astrophys. J., 583}, 334-357.
\refs Hartigan P., Strom K. M. and Strom S. E. (1994) {\em Astrophys. J., 427}, 961-977.
\refs Hartmann L., Calvet N., Gullbring E. and D'Alessio, P. (1998) {\em Astrophys. J., 495}, 385-400.
\refs Herbig G. H. (1948) {\em Publ. Astron. Soc. Pac.  60}, 256-267.
\refs Herbig G. H. (1989) in "The Formation and Evolution of
Planetary Systems", eds. H. Weaver L. Danly (Cambridge
U. Press Cambridge) p. 296-312.
\refs Hillenbrand L. A. and White R. J. (2004) {\em Astrophys. J., 604}, 741-757.
\refs Hillenbrand L. A., Strom S. E., Calvet N., Merrill K. M., Gatley I., Makidon R. B., Meyer M. R. and Skrutskie M. F. (1998), {\em Astron. J.,  116}, 1816-1841.
\refs Hodapp K. W., Bally J., Eisloffel J. and Davis C.J. (2005) {\em Astron. J.,  129}, 1580-1588.
\refs Holman M. J. and Wiegert P. A. (1999) {\em Astron. J.,  117}, 621-628.
\refs Jayawardhana R., Luhman K. L., D'Alessio, P. and Stauffer J. R. (2002) {\em Astrophys. J., 571}, L51-54.
\refs Jensen E. L. N., and Akeson R. L. (2003) {\em Astrophys. J., 584}, 875-881.
\refs Jensen E. L. N. and Mathieu R. D. (1997) {\em Astron. J.,  114}, 301-316.
\refs Jensen E. L. N., Mathieu R. D. and Fuller G. A. (1996) {\em Astrophys. J., 458}, 312-326.
\refs Jensen E. L. N.,  Mathieu R. D.,  Donar A. X. and Dullighan A. (2004) {\em Astrophys. J., 600}, 789-803.
\refs Joy A. H. and van Biesbroeck G.  (1944) {\em Publ. Astron. Soc. Pac.  56}, 123-124.
\refs Johnson H. L. (1966) {\em Ann. Rev. Astron. Astrophys.,  4}, 193-206.
\refs Koerner D. W., Jensen E. L. N., Cruz K. L., Guild T. B. and Gultekin K. (2000) {\em Astrophys. J., 533}, L37-40.
\refs Kenyon S.J. and Hartmann L. (1995) {\em Astrophys. J. Supp., 101},117-171.
\refs Knacke R. F., Strom K. M., Strom S. E., Young E., and Kunkel W. (1973) {\em Astrophys. J., 179}, 847-854.
\refs Krist, J. E., Stapelfeldt, K. R., and Watson, A.M. (2002)  {\em Astrophys. J., 570}, 785-792.
\refs Krist, J. E., Stapelfeldt, K. R., Golimowski, D. A., Ardila, D. R., Clampin, M., et al. (2005) {\em Astron. J., 130}, 2778-2787.
\refs Kuhn J. R., Potter D. and Parise B. (2001) {\em Astrophys. J.,, 553}, L189-191.
\refs Lay O. P, Carlstrom J. E., Hills R. E. and Phillips T. G. (1994) {\em Astrophys. J., 434}, L75-78.
\refs Leinert Ch. and Haas M. (1989) {\em Astrophys. J., 342}, L39-42.
\refs Leinert Ch., Zinnecker H., Weitzel N., Christou J., Ridgway S. T.,  Jameson R.,  Haas M. and Lenzen R. (1993) {\em Astron. Astrophys., 278}, 129-149. 
\refs Lissauer J. J., Quintana E. V., Chambers J. E., Duncan M. J. and Adams F. C. (2004) {\em Rev. Mex. Astron. Astrophys.,  22}, 99-103. 
\refs Lynden-Bell D. and Pringle J. E. (1974) {\em Mon. Not. R. Astron. Soc.,  168}, 603-637.
\refs Lubow S. H. and Ogilvie G. I. (2000) {\em Astrophys. J., 538}, L326-340. 
\refs Mart\'{\i}n E. L. (1998) {\em Astron. J.,  115}, 351 - 357.
\refs Maheswar G., Manoj P., and Bhatt H. C. (2002) {\em Astron. Astrophys.,, 387}, 1003-1012. 
\refs Malbet F., Lachaume R. and Monin J.-L. (2001) {\em Astron. Astrophys., 379}, 515-528
\refs Malbet F., Benisty, M., De Wit W. J., Kraus, S., Meilland, A. et al. (2005) {\em Astron. Astrophys., } in press (astro-ph/0510350). 
\refs Mathieu R. D. (1994) {\em Ann. Rev. Astron. Astrophys.,  32}, 465-530. 
\refs Mathieu R. D., Adams F. C. and Latham D. W. (1991) {\em Astron. J.,  101}, 2184-2198.
\refs Mathieu~R.~D., Adams~F.~.C, Fuller~G.~A., Jensen~E.~L.~N., Koerner~D.~W., and Sargent~A.~I. (1995) {\it Astron.~J., 109}, 2655-2669.
\refs Mathieu R.D., Stassun K., Basri G., Jensen, E.L.N., Johns-Krull, C.M., valenti, J.A. and Hartmann, L. W. (1997) {\em Astron. J.,  113},1841-1854.
\refs McCabe, C., Duchene, G., and Ghez, A.M. (2002) {\em Astrophys. J., 575}, 974-988.
\refs McCabe C., Ghez A. M., Prato, L., Duch\^ene G., Fisher R. S. and Telesco, C. (2006) {\em Astrophys. J., 636}, 932-951.
\refs McCaughrean M. J. and Rodmann J., 2006, in prep.
\refs M\'enard F., Monin J.-L., Angelucci F. and Rouan D. (1993) {\em Astrophys. J., 414}, L117-120.
\refs Mendoza V. and Eugenio E. (1966) {\em Astrophys. J., 143}, 1010.
\refs Mendoza V. and Eugenio, E. (1968) {\em Astrophys. J., 151}, 977.
\refs Monin J.-L. and Bouvier J. (2000) {\em Astron. Astrophys., 356}, L75-78.
\refs Monin J.-L., M\'enard F. and Duch\^ene G. (1998)  {\em Astron. Astrophys., 339}, 113-122.
\refs Monin J.-L., Menard F., and Peretto, N. (2001) {\em The Messenger 104}, 29-31.
\refs Monin J.-L., M\'enard F. and Peretto N. (2005) {\em Astron. Astrophys., 446}, 201-210.
\refs Moriwaki K. and Nakagawa Y. (2004) {\em Astrophys. J., 609}, 1065-1070.
\refs Mugrauer M., Neuhauser R., Seifahrt A., Mazeh T. and Guenther, E. (2005) {\em Astron. Astrophys., 440}, 1051-1060.
\refs Muzerolle J., Hartmann L. and Calvet N. (1998), {\em Astron. J.,  116}, 2965-2974.
\refs Najita J., Carr J. S. and Mathieu R. D. (2003) {\em Astrophys. J., 589}, 931-952.
\refs Nelson A.F. (2000) {\em Astrophys. J., 537}, L65-68.
\refs Ochi Y., Sugimoto, K. and  Hanawa T. (2005) {\em Astrophys. J., 623},922-939.
\refs Papaloizou J. C. B. and Pringle J. E. (1987) {\em Mon. Not. R. Astron. Soc.,  225}, 267-283.
\refs Papaloizou J. C. B., Terquem C. E. J. M. L. J. (1995) {\em Mon. Not. R. Astron. Soc.,  274}, 987-1001.
\refs Patience J., Ghez A. M., Reid I. N. and Matthews K. (2002) {\em Astron. J.,  123}, 1570-1602.
\refs Patience J., Akeson R. L., Jensen E. L. N. and  Sargent A. I. (2005) in {\em Protostars and Planets~V}, PLI contr No~1286, poster~\#8603.
\refs Potter D. E., Close L. M., Roddier F., Roddier C., 
	Graves J. E. and Northcott M. (2000) {\em Astrophys. J., 540}, 422-428.
\refs Prato, L. and Simon M. (1997) {\em Astrophys. J.,  474}, 455-463.
\refs Prato L., Ghez A. M., Pi\~{n}a R. K., Telesco C. M., Fisher R. S., Wizinowich P., Lai O., Acton D. S. and Stomski P. (2001) {\em Astrophys. J., 549}, 590-598.
\refs Prato L., Simon M., Mazeh T., Zucker S. and McLean I. S. (2002) {\em Astrophys. J., 579}, L99-102.
\refs Prato L., Greene T. P. and  Simon M. (2003) {\em Astrophys. J., 584}, 853-874.
\refs Quillen A. C., Blackman E. G., Frank A. and Varni\`ere P. (2004) {\em Astrophys. J., 612}, L137-140.
\refs Quintana E. V., Lissauer J. J., Adams F. C., Chambers J. E. and Duncan M.  J. (2005) In {\em Protostars and Planets~V}, LPI contribution No~1286, poster~\#8621.
\refs Reipurth B. (2000) {\em Astron. J., 120} 3177-3191.
\refs Reipurth B., Heathcote S., Roth M., Noriega-Crespo, A. and Raga A. C. (1993) {\em Astrophys. J., 408}, L49-52.
\refs Roddier C., Roddier F., Northcott M. J., Graves J. E., Jim K. et al. (1996) {\em Astrophys. J., 463}, 326-335.
\refs Rucinski S. M. (1985) {\em Astron. J.,  90}, 2321-2330.
\refs Rydgren A. E., Strom S. E., and Strom K. M. (1976) {\em Astrophys. J. Supp., 30}, 307-336.
\refs Scally A. and Clarke~C.~J. (2001) {\em Mon. Not. R. Astron. Soc.,  325}, 449-456.
\refs Simon M. and Guilloteau S. (1992) {\em Astrophys. J., 397}, L47-49.
\refs Simon M. and Prato  L. (1995) {\em Astrophys. J., 450}, 824-829.
\refs Simon M., Ghez A. M., Leinert Ch.,  Cassar L., Chen W. P., Howell R. R., Jameson R. F., Matthews K., Neugebauer G. and Richichi A. (1995) {\em Astrophys. J., 443}, 625-637.
\refs Simon M., Dutrey A. and Guilloteau S. (2000) {\em Astrophys. J., 545}, 1034-1043.
\refs Skrutskie M. F., Dutkevitch D., Strom S. E., Edwards S.,
Strom K. M., and Shure M. A. (1990) {\em Astron. J.,  99}, 1187-1195.
\refs Stapelfeldt K. R., Krist J. E., M\'enard F., Bouvier J., Padgett D. L. and Burrows C. J. (1998), {\em Astrophys. J., 502}, L65-69.
\refs Stapelfeldt K. R., M\'enard F., Watson A. M., Krist J. E., Dougados C., Padgett D. L. and Brandner W. (2003) {\em Astrophys. J., 589}, 410-418.
\refs Stassun K. G.,Mathieu R. D., Vrba F. J.,.
Mazeh T. and Henden A. (2001) {\em Astron. J.,  121}, 1003-1012.
\refs Stempels H. C. and Gahm G. F. (2004) {\em Astron. Astrophys., 421}, 1159-1168.
\refs Strom S. E. (1972) {\em Publ. Astron. Soc. Pac.  84}, 745-756.
\refs Strom K. M., Strom S. E., and Yost J. (1971) {\em Astrophys. J., 165}, 479-488.
\refs Strom K. M., Strom S. E., Breger M., Brooke A. L., Yost J., Grasdalen G. and Carrasco, L. 
(1972) {\em Astrophys. J., 173}, L65-70.
\refs Strom S. E., Strom K. M. and Grasdalen G. L. (1975) {\em Ann. Rev. Astron. Astrophys., 13}, 187-216.
\refs Strom S. E., Strom K. M., Grasdalen G. L., Capps R. W. and Thompson D. (1985) {\em Astron. J.,  90}, 2575-2580.
\refs Strom K. M., Strom S. E., Edwards S., Cabrit S. and  Skrutskie M. F. (1989) {\em Astron. J.,  97} 1451-1470.
\refs Takeuchi T., Clarke~C. J. and  Lin D. N. C. (2005) {\em Astrophys. J., 627}, 286-292.
\refs Tessier E., Bouvier J., Lacombe F. (1994) {\em Astron. Astrophys.,  283}, 827-834.
\refs Thebault P., Marzari F., Scholl, H., Turrini, D. and Barbieri M.,
2004, {\em Astron. Astrophys., 427}, 1097-1104 .
\refs Vicente S. M. and Alves J. (2005) {\em Astron. Astrophys., 441},195-205.
\refs White R. J. and Ghez A. M. (2001) {\em Astrophys. J., 556}, 265-295.
\refs Wolf S., Stecklum B. and Henning T. (2001) IAUS 200, 295-304.
 }
\end{document}